\documentclass[aps,preprint,floats,nofootinbib]{revtex4}
\usepackage{graphicx}
\usepackage{epsf}
\def\esph{E_{sph}} 
\def\w{\rm w} 

\def\lsim{\mathrel{\raise.3ex\hbox{$<$\kern-.75em\lower1ex\hbox{$\sim$}}}} 
\def\gsim{\mathrel{\raise.3ex\hbox{$>$\kern-.75em\lower1ex\hbox{$\sim$}}}}

\begin{document} 
 
\hfill$\vcenter{\hbox{\bf MADPH--03--1338}
                 \hbox{\bf hep--ph/0307120}}$

\vskip 0.5cm

\title {Effects of Electroweak Instantons\\  In High-Energy Neutrino Telescopes} 
\author{Tao Han and Dan Hooper} 
\address{   
Department of Physics, University of Wisconsin, 1150 University Avenue,   
Madison, WI 53706  }
\date{June, 2003}

\begin{abstract} 

We demonstrate that next generation high-energy neutrino telescopes 
may reveal the existence of interactions induced by standard model 
electroweak instantons. The energy spectrum, the angular distribution,
and the quark and lepton multiplicity of 
events in the detector each provide signatures which can indicate the 
presence of these interactions. High-energy neutrino telescopes may
be capable of searching for signals at energies far beyond the 
reach of the next generation colliders. 

\end{abstract} 

\pacs{PAC numbers:  13.15.+g, 95.85.Ry, 12.38.Lg  
\hspace{0.5cm} hep-ph/0307120,
\hspace{0.5cm} MADPH-03-1338}

\maketitle 
 
\section{Introduction}

Instantons are classical solutions of non-Abelian gauge theories 
in Euclidean space-time \cite{instant}, that represent tunneling 
transitions between topologically inequivalent vacua. 
The processes induced by electroweak instantons could be extremely 
interesting since they violate baryon+lepton number $(B+L)$ 
conservation \cite{thooft}.
The transition rate for the tunneling processes are exponentially suppressed 
at low energies compared to the the energy barrier separating the vacua, 
referred as the ``sphaleron" energy 
($\esph \approx \pi M_W/\alpha_{\w} \sim 8$ TeV) \cite{sphaleron}.
%typically of the order exp($-4\pi/\alpha_{\w}$).
However, at high temperatures \cite{highT}, or high energies \cite{highE},
the transition rate may be unsuppressed. It has thus been of great interest 
to explore the impact of the sphalerons on generating the baryon number
asymmetry of the Universe \cite{baryon}, 
and to ask if these effects can be observed in high-energy reactions 
such as at future colliders \cite{sphaleron,pheno}.

It is a challenge from a theoretical point of view to reliably estimate 
the two-body scattering rate for the instanton-induced processes near and
above $\esph$ \cite{Mattis:1991bj}.
In view of the close analogy of QCD, parton scattering amplitudes 
induced by electroweak instantons have been calculated using a perturbation 
method in the instanton background, with the saddle-point
approximation \cite{state}. It was argued that, due to the very
large pre-exponential factor in the cross section formula \cite{prefactor}
\begin{equation}
\sigma_0 \approx {(2\pi/\alpha_{\w})^{7/2}\over M_W^2}\approx 
5.3\times 10^{3}\ {\rm mb} ,
\label{pre}
\end{equation} 
the scattering cross section may be sufficiently enhanced at parton 
center-of-mass energies of about $E\ge 4\pi M_W/\alpha_{\w}\sim 30$ TeV, 
leading to possibly observable 
effects at a Very Large Hadron Collider (VLHC)  and
at future ultra-high energy cosmic ray experiments \cite{Fodor:2003bn}.
Recently, a new calculation based on a generalized semi-classical
approach has been performed to determine the exponent function \cite{rubakov}. 
Their numerical results demonstrated a severe exponential suppression, 
extended to energies as high as  $30\times \esph\sim 250$ TeV,
under the simplification of the $S$-wave dominance. 
Beside the phenomenological and observational excitement motivating searches 
for electroweak instanton effects, it would be of significant importance 
to obtain a quantitative description of such interactions and to test the 
different theoretical approaches, such as the two mentioned above. 
An initial comparison has just been made \cite{prefactor}
as we were completing this work.

We are approaching the era of neutrino astrophysics in which it will 
become possible to observe neutrinos of extra-galactic origin, at 
energies of PeV to EeV scales. Present experiment, such as 
AMANDA \cite{amanda}, Baikal \cite{baikal} and RICE \cite{rice} 
have successfully measured the atmospheric neutrino spectrum to 
energies up to $\sim 10 \,\rm{TeV}$ and have placed limits on any 
diffuse high-energy neutrino flux beyond this energy. IceCube \cite{icecube},
with a cubic kilometer effective 
volume, will be sensitive to charged and neutral current neutrino 
interactions from $\sim$TeV to the highest observed energies. Furthermore, 
planned experiments such as ANITA \cite{anita} and EUSO/OWL \cite{euso} 
will have even greater sensitivities to ultra-high energy neutrinos. 
For a review of high-energy neutrino astronomy, see Ref.~\cite{review}.

 In this Letter, we study in detail the  instanton or sphaleron-induced 
processes in high-energy neutrino telescopes, following some early
proposals \cite{duncan}. 
Neutrino telescopes can measure neutrino scattering 
cross sections at energies far greater than those accessible to colliders. 
At energies around $E_\nu\sim$100 TeV, the Earth becomes opaque to neutrinos
with the standard model (SM) cross sections. 
Taking advantage of this feature, comparing the flux of neutrinos
observed traveling through the Earth (up-going events) to the flux  
from above the detector (down-going), the neutrino-nucleon cross section 
can be inferred \cite{measure}. More generally, the zenith angle 
distribution of events in an underground detector can, given a sufficient flux, 
reveal this cross section.
Satellite-based experiments and air shower arrays cannot perform this task as 
effectively, as they are not designed to observe neutrino events over as great 
an angular distribution, or at as great a range of energies. A large volume, 
underground detector is ideal for this task. For this reason, 
we focus on the IceCube detector,  which is presently under construction 
and will be completed well before other post-LHC collider
 projects can be realized.

In our analysis, we consider both of the calculations for 
the cross section of instanton-induced neutrino-nucleon interactions described earlier.
For the calculation by Ringwald, we perform the
numerical evaluation by closely following the formalism in Ref.~\cite{state}.
As for the calculation by Bezrukov {\it et al.}~(BLRRT),
only the exponential function is given  \cite{rubakov}. 
We have thus assumed  a constant pre-factor for the cross 
section given by Eq.~(\ref{pre}), 
multiplied by the exponential function parameterized from \cite{rubakov}.
We note that this may have been a very crude estimate for the cross section
since the full pre-factor as in \cite{state} is energy-dependent and it may
decrease at higher energies. 

\begin{figure}[tb] 
%\vbox{\kern2.4in\special{ 
%psfile='cross.ps' 
%angle=90 
%hoffset=295 
%voffset=-51
%voffset=-78
%hscale=60
%vscale=60}}
\includegraphics[scale=0.6,angle=90]{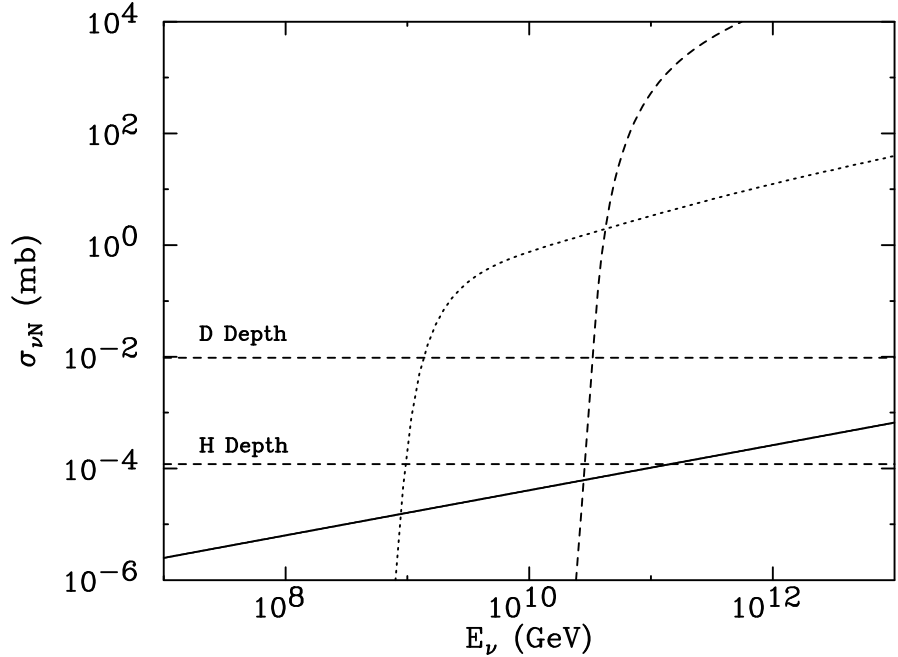}
\caption{Neutrino-nucleon cross sections via standard model electroweak 
instanton-induced processes. 
The dotted line represents the calculation of 
Ringwald. The dashed line represents the calculation of 
Bezrukov  {\it et al.}, using a 
pre-factor given in Eq.~(\ref{pre}).
The solid line is the standard model 
neutral+charged current prediction. Also shown as horizontal dashed 
lines are the cross sections which correspond to interaction lengths 
equal to the vertical down-going and horizontal depths of IceCube. 
}
\label{one} 
\end{figure} 

The cross sections predicted by these calculations are shown in Fig.~\ref{one}
by the dotted and dashed curves for the Ringwald's and BLRRT
calculations, respectively.  As expected, the cross section
grows sharply above the threshold, with a $\nu N$ center-of-mass energy about 30 TeV 
for the Ringwald's calculation, and as high as about 250 TeV for the 
BLRRT calculation.
For comparison, the standard model neutrino-nucleon total cross section 
of neutral and charged currents \cite{uhecr2} 
is also shown (solid line), where the three active flavors of neutrinos have
been included and the CTEQ-5 parton distribution 
functions \cite{cteq} are used.  Also illustrated as horizontal dashed 
lines are the cross sections which correspond to interaction lengths 
equal to the vertical down-going and horizontal depths of IceCube
(D Depth and H Depth).
It is in the range between these lines that distinctive features 
in the angular distribution of down-going events appear. 
For a neutrino  flux $\Phi$,   
the number of events $N_{\nu}$ observed as 
hadronic or electromagnetic showers in a neutrino detector of effective   
volume, $V$, is given by the convolution over energy of the quantity 
$V \times \Phi   \times n \times \sigma_{\nu}$. Here $n$ is the density of the   
target that interacts with a neutrino with cross section $\sigma_{\nu}$ 
to produce a shower. 

\begin{figure}[tb] 
%\vbox{\kern2.3in\special{ 
%psfile='WB-GRB-charm-largeD_energy.ps' 
%psfile='energy.ps' 
%angle=90 
%hoffset=-20 
%voffset=-85
%hoffset=290 
%voffset=-40
%hscale=41 
%vscale=45}} 
\includegraphics[scale=0.6,angle=90]{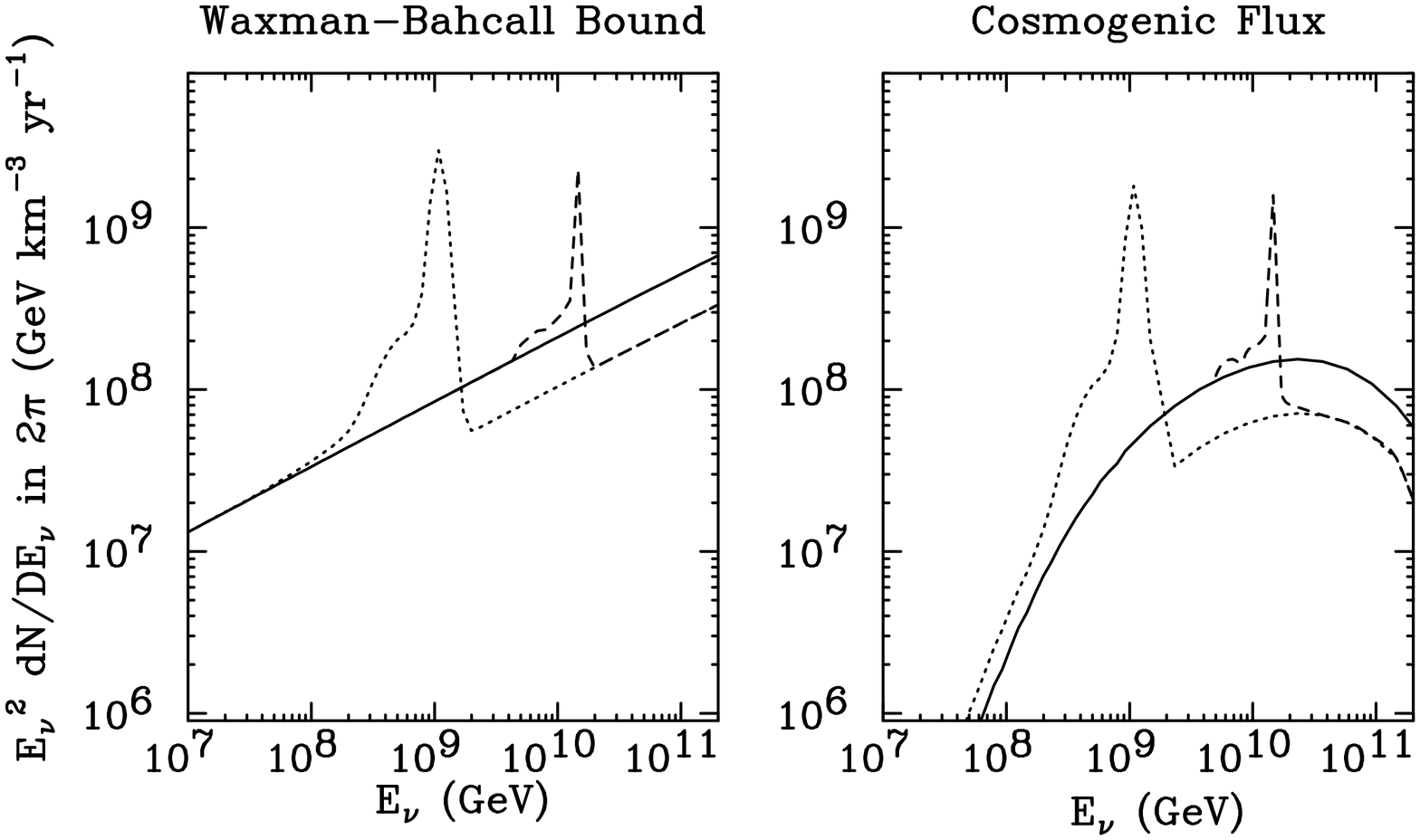}
\caption{Energy distribution of showers generated in neutrino-nucleon 
interactions via standard model electroweak instanton-induced processes,
with two assumed neutrino fluxes.  
The dotted line represents the calculation of 
Ringwald. The dashed line represents the calculation of 
Bezrukov  {\it et al.}, using a 
pre-factor given in Eq.~(\ref{pre}).
The solid line is the standard model neutral+charged current prediction.
}
\label{two} 
\end{figure} 

In Fig.~\ref{two}, the energy
spectrum of down-going shower events predicted in the IceCube 
experiment is shown. In the Ringwald's calculation (dotted line), 
as the center-of-mass neutrino-nucleon energy exceeds the sphaleron energy threshold
near 30 TeV, the number of events increases  dramatically 
above the standard model prediction. Even farther above this energy, however, 
more of the neutrinos are absorbed in the ice
before reaching the detector and the event 
rate is suppressed.  This drastic ``bump" structure in the spectrum 
indicates the sharply enhanced cross section at the sphaleron energy 
threshold position. The peak of this 
bump occurs at the associated neutrino energy and is mainly generated 
by charged current electron neutrino interactions. The ``shoulder'' 
slightly to the left of the bump  is from neutral and charged 
current interactions which generate showers less energetic than the incident neutrino.
These features occur at considerably higher energies for cross 
sections found using the 
calculation by Bezrukov {\it et.al.} \cite{rubakov}. In this approach, 
due to the exponential 
suppression of instanton-induced interactions well above the sphaleron energy, 
observations of these interactions will be considerably more difficult to 
study in neutrino telescopes, or colliders.

The left frame of Fig.~\ref{two} considers a flux of neutrinos equal to 
the upper bound found by Waxman and Bahcall \cite{wbbound}. This choice of 
flux represents neutrinos from compact engines, such as gamma ray bursts or 
hadronic blazars. The limit of Waxman and Bahcall is somewhat controversial 
and it has been argued that larger high-energy neutrino fluxes may be possible 
from such sources \cite{mpr}. The right frame considers the neutrino flux 
from the interactions of ultra-high energy protons with the cosmic microwave 
background, called the cosmogenic neutrino flux \cite{cosmogenic}. 
We have used the cosmogenic flux as calculated in Ref.~\cite{cosmogenic2}. 
The cosmogenic neutrino flux can be reliably calculated from the observed 
flux of ultra-high energy protons. Therefore, the choice of this flux is 
quite conservative. It is interesting to note that at EeV energies, 
the Waxman-Bahcall bound is not far above the conservative
cosmogenic prediction.
For this reason, in our discussion the detected neutrino flux plays a secondary role.

\begin{figure}[tb] 
%\vbox{\kern2.4in\special{ 
%psfile='angleEeV.ps' 
%angle=90 
%hoffset=270 
%hoffset=-10 
%voffset=-60 
%hoffset=280 
%voffset=-36
%hoffset=290 
%voffset=-53
%hscale=40
%vscale=38}}
\includegraphics[scale=0.56,angle=90]{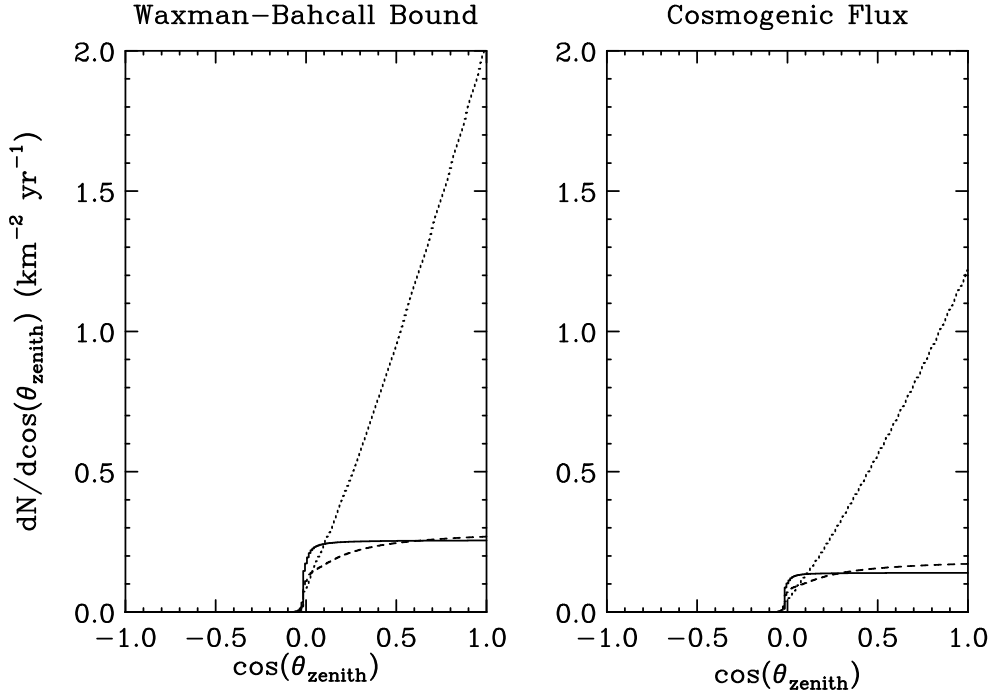} 
\caption{Zenith angle distribution of showers generated in neutrino-nucleon 
interactions via standard model electroweak instanton-induced processes,
with two assumed neutrino fluxes. 
The dotted line represents the calculation of 
Ringwald. The dashed line represents the calculation of 
Bezrukov  {\it et al.}, using a 
pre-factor given in Eq.~(\ref{pre}).
The solid line is the standard model 
neutral+charged current prediction.
A 1 EeV energy threshold for the observed showers has been imposed. 
Up-going events correspond to $-1<\cos\theta_{\rm zenith}<0$, whereas 
down-going events correspond to $0<\cos\theta_{\rm zenith}<1$. 
}  
\label{three} 
\end{figure}

Figure~\ref{three} shows the zenith angle distribution of showers above 1 EeV 
in a kilometer-scale detector. 
Up-going events correspond to $-1<\cos\theta_{\rm zenith}<0$, whereas 
down-going events correspond to $0<\cos\theta_{\rm zenith}<1$. 
For standard model interactions, the distribution  (solid curve) 
is nearly flat for down-going events, and essentially no up-going events occur 
due to very efficient neutrino absorption by the Earth at these energies. 
For models with larger cross sections from instanton-induced interactions, 
vertical down-going events become more frequent, producing
more events near $\cos\theta_{\rm zenith}\sim 1$. 
At zenith angles near the horizon, $\cos\theta_{\rm zenith}\sim 0$, 
more of the neutrinos are absorbed and the rate can be suppressed.
We have chosen the shower threshold of 1 EeV to optimize the angular
effects for the case of Ringwald's calculation.

Another interesting characteristic feature of instanton-induced processes
is the large multiplicity of final state particles and the violation of $B+L$.
The basic operators involving quark and lepton fields are of the 
form $\langle (qqq\ell)^{n_g} \rangle$ \cite{thooft}, where $n_g=3$ is
the number of fermion generations. It has been argued that the processes
involving multiple gauge bosons and Higgs bosons, 
like $\langle (qqq\ell)^{n_g} W^n H^m \rangle$, can be significantly
enhanced \cite{highE}. A typical neutrino-induced event could thus be
\begin{equation}
\nu_e u \to \bar d \bar d\  
+ \ \bar c \bar c\bar s \mu^+
+ \ \bar t \bar t\bar b \tau^+\
+ nW + mH.
\end{equation}
With both quarks and leptons of all three generations involved simultaneously
at the primary production, this type of events should look quite unique. 
It is, however,
difficult to predict how the events would exactly look like in the IceCube detector
given the fact that the particles are highly collimated and will be challenging 
to separate.

 \begin{table}[tb]
 \begin{tabular}{c c c c c}
 \hline \hline
 &~~&~Waxman-Bahcall~&~Cosmogenic~&~\\
 \hline \hline
 &~0.4 EeV and above~&~~&~~&~\\
 \hline 
&CC+NC&0.24~&~0.12~&~\\
&Ringwald&1.3~&~0.72~&~\\
& BLRRT&0.26~&~0.14~&~\\
 \hline 
 &~$0.4-2.0$ EeV~&~~&~~&~\\
 \hline
&CC+NC&0.15~&~0.070~&~\\
& Ringwald&1.2~&~0.70~&~\\
& BLRRT&0.15~&~0.070~&~\\
 \hline \hline
 &~$10-20$ EeV~&~~&~~&~\\
 \hline 
&CC+NC&0.013~&~0.0071~&~\\
& Ringwald&0.0063~&~0.0033~&~\\
& BLRRT&0.034~&~0.023~& ~\\
 \hline \hline
 \end{tabular}
\caption{
Event rates (showers) in a high-energy neutrino telescope per year, 
per cubic kilometer of effective volume. 
Rates are shown for the standard model 
charged plus neutral current (CC+NC), 
as well as for the calculations by Ringwald and by BLRRT
(including CC+NC). 
Rates for three (shower) energy ranges are shown. 
These were selected to illustrate the features in the energy spectrum associated 
with each approach.}
\label{four} 
\end{table}

Shown in Table \ref{four} are the event rates predicted in a kilometer-scale 
high-energy neutrino telescope, such as IceCube. 
Rates are shown for the standard model 
charged plus neutral current (CC+NC), 
as well as for the calculations by Ringwald and by BLRRT (including CC+NC).
Rates are shown for three (shower) energy ranges, chosen to illustrate the 
features in the energy spectrum associated with each approach (see Fig.~\ref{two}). 
For the Ringwald's calculation, the event rate predicted 
for the cosmogenic neutrino flux is about $0.7$ per year 
in the narrow range of 0.4 to 2 EeV, 
a factor of ten above the SM (CC+NC) prediction. 
This rate is somewhat higher (1.2) for the flux given by the
Waxman-Bahcall bound. Either of these cases provide a possible signature for 
observations over several years.
The predictions of BLRRT are considerably more difficult to test, 
however. In the peaked region of the energy spectrum, at 10 to 20 EeV, 
only on the order of 0.03 events per year are expected. 
Although this is much larger than the SM (CC+NC)
prediction, it will be very challenging for 
kilometer-scale instruments to observe this signature. 
An order of magnitude larger detector in effective volume
would be needed to probe such a scenario.

It has been proposed that ultra-high energy neutrinos undergoing instanton-induced 
interactions may have generated many of the observed cosmic ray events above the 
GZK cutoff \cite{Fodor:2003bn}. Although it would be extremely interesting to 
establish this interpretation,
it will be difficult to determine with confidence that this is the case with 
air shower experiments.  Neutrinos with mb scale cross sections 
have similar experimental signatures to protons and, therefore, will be 
difficult to distinguish.

Future satellite-based cosmic ray experiments, such as EUSO/OWL \cite{euso}, 
may be able to observe similar numbers of ultra-high energy neutrino events 
compared to IceCube \cite{halzen20}. Much like ground based cosmic ray experiments, 
EUSO/OWL does not have the phenomenological advantages of a deeply buried 
neutrino telescope described in this paper. Such experiments will have a 
more difficult time identifying neutrino-nucleon cross section enhancements. 

In summary, the SM electroweak instantons may provide observable 
signatures in kilometer-scale high-energy neutrino telescopes. 
The large deviations of the neutrino-nucleon cross 
section from the charged and neutral current prediction found in these 
calculations provide distinctive features 
in both the spectrum and angular distribution of events in the detector.
These variations reflect both the increased probability of a neutrino 
interacting in the detector and the increased probability of being 
absorbed in the Earth.
Another qualitative feature would be presented by the particle content 
of the events (all three generations of quarks and leptons with $B+L$ violation). 

The calculation of Ringwald's should be unambiguously tested by IceCube
before post-LHC colliders can be realized. 
The prediction of BLRRT will be considerably more difficult to probe
due to the suppression of the rate in the relevant energy region. 

{\it Acknowledgments}: This work was supported in part by a DOE grant No.  
DE-FG02-95ER40896  and in part by the Wisconsin Alumni Research Foundation.

\end{document}